# Ozone and OH-induced oxidation of monoterpenes: Changes in the thermal properties of secondary organic aerosol (SOA)

*Ågot K. Watne[1], Jonathan Westerlund[1], Åsa M. Hallquist[2], William H. Brune[3] and Mattias Hallquist[1,*]*

[1]Atmospheric Science, Department of Chemistry and Molecular Biology, University of Gothenburg, SE-41296 Gothenburg, Sweden

[2]IVL Swedish Environmental Institute, PO Box 5302, SE-400 14 Gothenburg, Sweden

[3]Department of Meteorology and Atmospheric Science, Pennsylvania State University, University Park, PA, USA

[*]Corresponding author; hallq@chem.gu.se

**Abstract**

The behaviour of secondary organic aerosols (SOA) in the atmosphere is highly dependent on their thermal properties. Here we investigate the volatility of SOA formed from α-pinene, β-pinene and limonene upon ozone- and OH-induced oxidation, and the effect of OH-induced ageing on the initially produced SOA. For all three terpenes, the ozone-induced SOA was less volatile than the OH-induced SOA. The thermal properties of the SOA were described using three parameters extracted from the volatility measurements: the temperature at which 50% of the volume has evaporated ($T_{VFR0.5}$), which is used as a general volatility indicator; a slope factor ($S_{VFR}$), which describes the volatility distribution; and $T_{VFR0.1}$, which measures the volatility of the least volatile particle fraction. Limonene-derived SOA generally had higher $T_{VFR0.5}$ values and shallower slopes than SOA derived from α- and β-pinene. This was especially true for the ozone-induced SOA, partially because the ozonolysis of limonene has a strong tendency to cause SOA formation and to produce extremely low volatility VOCs (ELVOCs). Ageing by OH exposure did not reduce $T_{VFR0.5}$ for any of the studied terpenes but did increase the breadth of the volatility distribution by increasing the aerosols' heterogeneity and contents of substances with different vapour pressures, also leading to increases in $T_{VFR0.1}$. This stands in contrast to previously reported results from smog chamber experiments, in which $T_{VFR0.5}$ always increased with ageing. These results demonstrate that there are two opposing processes that influence the evolution of SOAs' thermal properties as they age, and that results from both flow reactors and static chambers are needed to fully understand the temporal evolution of atmospheric SOA' thermal properties.

**Highlights:**

- Thermal properties of SOA from VOCs vary in their responses to oxidation processes

- Limonene-derived SOAs' thermal properties differ sharply from those for α- and β-pinene

- OH-induced oxidation yielded more volatile SOA than ozonolysis

- SOA formed by ozonolysis evaporated over a narrower temperature interval

- The observed responses to OH-induced ageing explain some unresolved observations

## 1. Introduction

Secondary organic aerosols (SOA) formed by the oxidation of organic trace gases and subsequent gas to particle conversion have important effects on atmospheric particulate matter because they affect its general composition and also its physicochemical properties. The key reactants that initiate SOA formation via the oxidation of VOCs in the troposphere are hydroxyl (OH) radicals and ozone ($O_3$) and the nitrate radical (Hallquist et al., 2009); globally, the vast majority of SOA formation is from the oxidation of biogenic monoterpenes such as α-pinene, β-pinene and limonene . The products generated from the oxidation of monoterpenes by these agents, and the composition of the subsequently generated SOA, depend on both the oxidant and the molecular structure of the monoterpene (Emanuelsson et al., 2013; Jonsson, Hallquist, & Saathoff, 2007; Zhao et al., 2015). Moreover, it was recently suggested that extremely low volatility organic compounds (ELVOC) formed during terpene oxidation play a crucial role in the formation of new particles (Ehn et al., 2014).

Atmospheric SOA can change over time as they undergo photochemical ageing. Ageing normally reduces particles' volatility because it involves the oxidation of their constituents (Donahue et al., 2012; Emanuelsson et al., 2014; Kalberer et al., 2004). In contrast, it has been speculated that under some circumstances, extensive gas-phase oxidation can increase volatility by promoting the condensation of more volatile compounds than those present in the initial aerosol (Salo et al., 2011; Tritscher et al., 2011).

Previous studies on the photochemical ageing of SOA have primarily been conducted using Potential Aerosol Mass (PAM) chambers (Hamilton, Lewis, Reynolds, Carpenter, & Lubben, 2006; Kang, Root, Toohey, & Brune, 2007). A PAM chamber is a photochemical reactor that can generate high concentrations of hydroxyl radicals to mimic OH-induced ageing of aerosols over periods of several days in the atmosphere. Among other things, the PAM chamber has been used to investigate the effects of photochemical ageing on cloud condensation nuclei activation (Lambe et al., 2011; Lambe et al., 2015), as well as the optical properties (Lambe et al., 2013), chemistry (Chhabra et al., 2015; Lambe et al., 2015), and yield (Bruns et al., 2015) of SOA, and to conduct ambient measurements (Ortega et al., 2015; Palm et al., 2015). However, to our knowledge, there are currently no published studies on the influence of photochemical ageing on the thermal properties of SOA.

Thermal properties such as the volatility distribution of a complex aerosol can be determined by using a volatility tandem differential mobility analyser (VTDMA) to measure the volume fraction remaining (VFR) as a function of the evaporation temperature. The TDMA technique was first described by Rader and McMurry (1986), and has since been used in laboratory studies to determine aerosols' physical properties such as saturation vapour pressures (Salo, Jonsson, Andersson, & Hallquist, 2010) and to investigate changes in the thermal properties of aerosols due to changes in their chemical composition (Emanuelsson et al., 2013; Emanuelsson, Watne, Lutz, Ljungström, & Hallquist, 2013; Jonsson et al., 2007; Salo et al., 2011). In the field, it has been used to conduct ambient measurements of general aerosol volatility under different atmospheric conditions (Wehner et al., 2009). Emanuelsson et al. (2013) used data from a VTDMA to describe SOA volatility in terms of two general properties: the overall volatility and its distribution. These properties were quantified by fitting a sigmoidal curve to the evaporation data from the VTDMA system. The overall volatility was defined in terms of the temperature at which 50% of the aerosol had evaporated, $T_{VFR0.5}$, for a given residence time in the heated section of the oven. The volatility distribution was defined by the steepness of the VFR slope, $S_{VFR}$. Because the VFR always decreases with the temperature, the slope is always negative. An aerosol with a more heterogeneous mixture of constituents (each having different vapour pressures) will have a broader volatility distribution, i.e. a shallower slope. Therefore, the value of $S_{VFR}$ approaches 0 as the volatility distribution broadens. More homogeneous aerosols have narrower volatility distributions, so a greater fraction of the aerosol is evaporated over a smaller temperature interval. That is to say, the slope becomes steeper and approaches $-\infty$ as the distribution becomes progressively more narrow.

To better understand the processes involved in the formation and ageing of SOA, and the effect of these processes on the aerosol's thermal properties, it is necessary to perform experiments under controlled conditions, for example using flow reactors and smog chambers. Here we present results from oxidation and ageing studies performed using combinations of two flow reactors. The aim of this work was to study the thermal properties of SOA formed by OH- and $O_3$-induced oxidation of three terpenes (α-pinene, β-pinene and limonene), and to evaluate the effects of subsequent OH-induced ageing on these SOA.

## 2. Experimental details

### 2.1 Flow reactors and experiments

An apparatus consisting of two flow reactors connected in series was used to establish selective oxidation conditions for three monoterpenes: α-pinene, β-pinene and limonene, see Figure 1. The first reactor was the Göteborg Flow Reactor for Oxidation Studies at low Temperature (G-FROST), a laminar flow reactor that has previously been used to study the ozonolysis of VOCs (Emanuelsson et al., 2013; Å M. Jonsson, M. Hallquist, & E. Ljungström, 2008; Jonsson, Hallquist, & Ljungström, 2008; Jonsson et al., 2007; Pathak et al., 2012). The second reactor was the PAM chamber which is a flow-through reactor capable of maintaining high oxidant concentrations; this ability was used to simulate several days of photo-oxidation over a period of only a few minutes in the laboratory (Kang et al., 2007). The two flow reactors were used together to determine how the thermal characteristics of SOA are affected by different oxidation environments. The total flow through the G-FROST reactor was 1.6 LPM, and 0.6 LPM of the centre fraction was sampled. The sampled flow was then diluted ten-fold with a mixture of humidified synthetic air and ozone to yield a total flow of 6 LPM into the PAM chamber.

Four different oxidation conditions were applied: OH-dominated oxidation (denoted OH); ozone-dominated oxidation with some influence from OH radicals produced during terpene ozonolysis (denoted MIX); ozonolysis with an OH scavenger to suppress the influence of OH (denoted O3); and sequential oxidation involving an initial ozonolysis of the terpenes as in the MIX case, followed by OH oxidation (denoted SEQ). Details of the experimental settings used to establish and maintain these conditions are provided in Table 1.

The terpenes and the OH-scavenger were introduced into the flow reactors via a dry bulk flow of synthetic air. Terpene concentrations were measured by sampling using adsorbent tubes followed by GC-FID analysis. In the absence of oxidation, the concentrations of the terpenes exiting the final PAM chamber were 6.6, 4.3 and $3.9 \times 10^{11}$ molecules $cm^{-3}$ (27, 18, and 16 ppb) for α-pinene, β-pinene and limonene, respectively. Because the reacting mixture is diluted before entering the PAM, the terpene concentration during the oxidation in the G-FROST reactor is ten times higher than that in the PAM. $O_3$ for use in the experiments was generated in the gas streams entering the flow reactors by irradiating $O_2$ with mercury lamps (BHK lamp, SCD 82-9304-03, λ=185 nm). The ozone concentration was monitored using a 49C UV photometric $O_3$ analyser (Thermo Environmental Instruments, Inc.). In the absence of terpenes, the concentration of $O_3$ was set to $7.4 \times 10^{12}$ molecules $cm^{-3}$ (300 ppb) in the PAM chamber. In certain experiments, 2-butanol was added to scavenge any produced OH radicals.

In the PAM chamber, OH radicals were produced by irradiating the existing $O_3$ using two mercury pen-ray lamps (λ=254 nm) to yield excited oxygen atoms that subsequently reacted with water. The rate of OH production was varied by changing the intensity of irradiation, and the corresponding OH exposure was determined indirectly in separate experiments by measuring the OH-induced decay of a known added concentration of $SO_2$ (Lambe et al., 2011; Peng et al., 2015). In the PAM chamber, the OH-exposure varied between $0.8 - 4.5 \times 10^{11}$ molecules $cm^{-3}$ $s^{-1}$. The RH in the PAM chamber was maintained at 30 % by diluting the dry flow (< 5 % RH) from the G-FROST reactor with humidified air. Both flow reactors were kept at room temperature (21.7 ± 1.7 ºC) during the experiments. Before each experiment, the PAM chamber was heated to 60 ºC with concurrent OH production so as to reduce the rate of particle formation due to OH oxidation of clean air (i.e. air without added terpenes) to less than 0.1 μg $m^{-3}$.

## 2.3 Particle monitoring and analysis

The particles produced in the experiments were characterised on the basis of their size and thermal properties. Particle number size distributions (10-300 nm) were measured with a scanning mobility particle sizer (SMPS 3936L75, TSI) operated in recirculating mode. The number size distributions were then used to estimate the aerosols' masses by assuming that the particles were spherical and had a density of 1.4 g cm$^{-3}$ (Hallquist et al., 2009). Thermal properties were determined using a volatility tandem differential mobility analyser (VTDMA). This instrument has been described in detail elsewhere (Jonsson et al., 2007; Salo et al., 2010); in brief, it has three main components. The first is a DMA, which is used to filter out particles outside a certain size range so as to establish a monodisperse aerosol; in these experiments, particles with a diameter of 87 nm were selected. The second is an oven unit in which the monodisperse aerosol is evaporated at a set temperature and then passed through a charcoal denuder in a short cooling section. The third is an SMPS unit that is used to measure the residual particle size distribution after evaporation at the specified oven temperature. The volume fraction remaining (VFR) at each evaporation temperature was determined from the median electrical mobility diameter of the monodisperse aerosol, assuming spherical particles, i.e. VFR = $[D_f/D_i]^3$, where $D_f$ is the final and $D_i$ is the initial median diameter. Figure 2 shows an example of the full size distribution before selection with the initial DMA and three resulting size distributions after selection and evaporation at respectively temperature. The diameter $D_f$ was determined as the number median peak diameter, the diameter with 50% of number of particles below and 50% above. The VTDMA sampling flow was 0.3 SLPM, giving a median residence time of 1.8 s in the heated part of the oven for a fully developed laminar flow. The VFR data at different evaporation temperatures were evaluated further using the method of Emanuelsson et al. (2013). A sigmoidal curve was fitted to the full range of VFR data as a function of the evaporation temperature (see Equation 1).

$$VFR_T = VFR_{min} + \frac{(VFR_{max} - VFR_{min})}{1 + \left(\frac{T_{position}}{T}\right)^{S_{VFR}}}. \qquad (1)$$

The exponent $S_{VFR}$ and the parameter $T_{position}$ determine the steepness ("slope factor") and mid-point of the symmetric sigmoidal curve, respectively. $VFR_{max}$ and $VFR_{min}$ represent the highest and lowest VFRs, respectively. To obtain the best fit to the data, $VFR_{max}$ and $VFR_{min}$ were not restricted *a priori*. From this equation, the temperature at which the VFR is 0.5 ($T_{VFR0.5}$) can be readily calculated; this quantity was used as a general volatility indicator that is inversely correlated with the volatility of the aerosol particles. The slope factor $S_{VFR}$ is a measure of the distribution of the aerosol's volatility, and is always negative because the VFR always decreased with increasing temperature. An aerosol with a broader volatility distribution has more constituents with different vapour pressures and thus a shallower slope; the value of $S_{VFR}$ approaches zero as the distribution becomes broader, as discussed in the introduction section.

In addition, the temperature corresponding to a VFR of 0.1 ($T_{VFR0.1}$) was calculated. This quantity reflects the volatility of the species responsible for the tail of the sigmoidal curve, i.e. the least volatile components of the SOA. As such, it may be a useful indicator for the presence of so-called extremely low volatility organic carbons (ELVOC).

## 3. Results and discussion

### 3.1 Mass and number

The experiments that were conducted and the key findings obtained are summarized in Table 1. As expected (and in keeping with previously reported findings), the oxidation conditions and extent of oxidation affected the total number and mass of particles that were formed (Emanuelsson et al., 2013; Jonsson et al., 2008; Jonsson et al., 2007; Pathak et al., 2012). While the focus of this work is on the thermal properties and ageing of the SOA, concurring measured masses and particle numbers support our previously established understanding of the studied systems.

For all three terpenes, the addition of an OH scavenger during the ozonolysis experiments reduced the total number and mass of particles that were formed. In fact, in the experiments conducted with β-pinene, no detectable SOA formation occurred when the OH scavenger was present. The decrease in particle mass when the scavenger was present was more pronounced for α-pinene (-52 %) than for limonene (-12 %); the corresponding decreases in particle number concentrations were 88 % and 43 %, respectively. These findings indicate that for α- and β-pinene the relative influence of OH chemistry compared to $O_3$ chemistry on SOA formation is larger than for limonene. It is interpreted as ozonolysis of limonene is very efficient in particle production so any additional OH-induced chemistry will have limited effects while ozonolysis of α- and β-pinene is not as effective and the additional OH chemistry will enhance particle production.

## 3.2 Thermal properties

Figure 3 shows the VFR as a function of the evaporation temperature and the corresponding fitted sigmoidal curves for the three terpenes and the different oxidation conditions. As can be seen from Figure 3, the least volatile aerosols are those formed by ozonolysis in presence of an OH-scavenger and the most volatile aerosols are formed when the terpenes are oxidised mainly by OH-radicals. Ozonolysis without an OH scavenger (MIX) yields aerosols of intermediate volatility, showing that volatility increases with the extent of OH-based oxidation.

To better characterize the effects of the oxidation conditions and the different precursor terpenes on the thermal properties of the SOA, three key variables were considered: $T_{VFR0.5}$, $T_{VFR0.1}$ and $S_{VFR}$. As shown in Table 1, the calculated $T_{VFR0.5}$ values are consistent with the general volatility trends presented in Figure 3. The ozonolysis of limonene (L-O3) gave the highest $T_{VFR0.5}$ (386.9 ± 0.1 K) and the oxidation of α-pinene with OH radicals (α-OH) gave the lowest (342.4 ± 0.2). Additionally, the $T_{VFR0.5}$ values for the limonene-derived SOA formed by ozonolysis in the presence (L-O3) and absence (L-MIX) of the OH scavenger were 12 and 14 K higher, respectively, than the corresponding values for α-pinene SOA (α-O3 and α-MIX, respectively). This is consistent with previously reported findings (Jonsson et al., 2007; Kolesar, Chen, Johnson, & Cappa, 2015; B.-H. Lee, Pierce, Engelhart, & Pandis, 2011) and the fact that unlike the pinenes, limonene has two double bonds that can undergo oxidation. In keeping with the results of Lee et al. (2010) and Kolesar et al. (2015), we observed no major differences in $T_{VFR0.5}$ between SOA derived from α- and β-pinene by ozonolysis in the absence of an OH scavenger (α and β-MIX). The $T_{VFR0.5}$ for the SOA derived from α-pinene by OH oxidation (α-OH) was somewhat lower (5 K) than that for the corresponding β-pinene-derived SOA (β-OH). The $T_{VFR0.5}$ values for limonene- and α-pinene-derived SOA formed by ozonolysis alone (O3) were 30 and 40K higher, respectively, than those for the corresponding SOA formed by OH oxidation (OH). The most pronounced difference between the $T_{VFR0.5}$ values for limonene- and α-pinene-derived SOA was 14K, which occurred for SOA formed by ozonolysis alone; this difference was smallest (4K) for SOA formed by OH oxidation.

The trends for $T_{VFR0.1}$ were generally similar to those for $T_{VFR0.5}$ but were more pronounced; for example, the $T_{VFR0.1}$ for the three terpenes showed very clearly that limonene behaved differently to α- and β-pinene. The highest $T_{VFR0.1}$ value was observed for L-O3 (449 K); this value is 34 K higher than that for α-O3. For comparative purposes, as noted in the preceding section, the $T_{VFR0.5}$ values for the same pair of experiments differed by only 14 K. Also for the mixed cases a higher $T_{VFR0.1}$ is found for limonene compared to α- and β-pinene. These results are consistent with the previously reported findings that limonene oxidation produces higher quantities of ELVOC than does that of α- or β-pinene (Ehn et al 2014, Jokinen et al., 2015). The difference in $T_{VFR0.1}$ between limonene-derived SOA formed by ozonolysis (O3) and OH oxidation (OH) was 45 K, compared to a difference of 40 K in their $T_{VFR0.5}$ values. However, for α-pinene the corresponding difference in $T_{VFR0.1}$ between SOA formed by ozonolysis and OH oxidation was only 15 K, which is significantly lower than the difference in $T_{VFR0.5}$ (30 K).

The slope variable $S_{VFR}$ describes the distribution of the vapour pressures of the aerosol's constituents. The lowest and highest $S_{VFR}$ values were obtained for the α-O3 (-20.8) and L-MIX (-11.3) SOA, respectively. Generally, the $S_{VFR}$ values for SOA formed in the presence of the OH-scavenger (i.e. under O3 conditions) were 4-18% lower than those for SOA formed without the scavenger (MIX). This is consistent with the expectation that the presence of both OH and ozone would yield a more diverse product mixture. The $S_{VFR}$ values for SOA derived from α- and β-pinene by ozonolysis are up to 16 % lower than those for SOA formed by OH oxidation. In contrast, the slope for limonene-derived SOA formed by ozonolysis in the absence of a scavenger (L-MIX) is around 30 % greater than that for the limonene-derived SOA formed by OH oxidation (L-OH). Furthermore, the $S_{VFR}$ for the limonene-derived SOA formed by ozonolysis (L-mix) is greater than those for the corresponding α- and β-pinene-derived SOA (α- and β-MIX) by 50% and 35%, respectively. This can again be explained by the fact that limonene has two double bonds and can thus give rise to a more complex product mixture than the pinenes, which have only one double bond. Overall, the results for all three volatility parameters ($S_{VFR}$, $T_{VFR0.1}$ and $T_{VFR0.5}$) suggest that the oxidation of limonene produces a very different volatility distribution to that for α- or β-pinene, in keeping with the observed differences in particle masses and numbers.

### 3.3 OH-induced ageing

The effect of OH-induced ageing on the SOA was studied by increasing the PAM OH exposure. In general, the mass of particles and particle numbers increased with the PAM OH exposure for all SOA, regardless of the conditions under which they were initially formed. However, the magnitude of these increases depended on the initial oxidation conditions and the initial terpene. As shown in Figure 4b, the ageing of α- and β-pinene SOA formed by ozonolysis caused sharp increases in number concentrations (124% and 514 %, respectively). Conversely, ageing of the corresponding limonene SOA caused only a very modest increase in number concentration (7%). However, it should be noted that the initial aerosol formed by limonene ozonolysis had a high particle number due to more efficient nucleation, which would reduce the scope for new particle formation upon further oxidation by OH. In contrast, the initial ozonolysis of β-pinene produced far fewer particles (when an OH scavenger was present, the number of particles was below the limit of detection). Therefore, the ageing of β-pinene SOA with OH induces substantial new particle nucleation, leading to the large observed increase in particle numbers.

PAM OH exposure also caused substantial increases in mass concentration (Fig. 4a). The mass increases observed in the sequential oxidation experiments ranged from 21- 77 % depending on the precursor terpene. Interestingly, as the PAM OH exposure increased, the particle mass rose to a peak and then declined slightly for SOA derived from limonene and α-pinene. Similar results have been reported for other SOA when subjected to comparable OH-exposures (~$10^{11}$ molecules cm$^{-3}$ s$^{-1}$) (Kang, Toohey, & Brune, 2011; Lambe et al., 2014; Lambe et al., 2012; Ortega et al., 2015; Palm et al., 2015; Tkacik et al., 2014). However, no such maximum was observed for β-pinene in the sequential oxidation experiments or any of the OH-dominated experiments. Apparently, β-pinene oxidation produces a pool of compounds in the gas phase that can contribute to particle mass increases even after extensive ageing. This can be plausible since it e.g. is clear that the exocyclic double bond in β-pinene could more easily lead to fragmentation when oxidised with ozone, creating one C9 and one C1 entity in comparison to α-pinene that by an analogue C-C cleavage at the position of the double bond will only give one entity (a C10). See the work of Emanuelsson et al. (2013) for a detailed discussion on the oxidation of β-pinene. Therefore, oxidation of the β-pinene producing C9 entities will give a greater proportion of short carbon chain products with comparatively high vapour pressures than would be formed from α-pinene, necessitating a greater degree of oxidation before they can contribute to the aerosol mass.

In the present study the PAM chamber was used to study how ageing by OH oxidation alters the thermal properties of SOA formed by ozonolysis. The results obtained in these experiments can be compared to the effects of increasing OH exposure without pre-forming an aerosol by ozonolysis, i.e. the ageing of an OH-induced SOA. Figure 5 shows the changes in $T_{VFR0.5}$, $T_{VFR0.1}$ and $S_{VFR}$ as a function of PAM OH-exposure for systems of both kinds. Similar trends are visible in both cases, and as shown in Figure 5a, increasing the PAM OH-exposure caused no appreciable change or only a slight decrease in $T_{VFR0.5}$. This stands in contrast to observations obtained with larger static chambers, where the volatility generally decreased with increasing OH exposure (Donahue et al., 2012; Emanuelsson et al., 2014; Salo et al., 2011; Tritscher et al., 2011). Emanuelsson et al (2013) quantified the change in $T_{VFR0.5}$ as a function of OH exposure using an ensemble of BVOC photooxidation experiments done in the large Julich smog chamber SAPHIR; they observed an increase of 0.3±0.1 % in $T_{VFR0.5}$ per hour at an OH concentration of $1 \times 10^6$ molecules $cm^{-3}$. The difference between these results and those of the present PAM chamber experiments in which no such increase was observed can be explained by the fact that the aerosol in the PAM chamber is continuously replenished and thus always reflects a freshly oxidised mixture at a given OH exposure, whereas static chambers by definition enable a second dimension of ageing, allowing time for other processes (e.g. wall uptake, chemical reactions that occur over longer time scales, and, in the case of SAPHIR, a continuous dilution process) to influence the results. The results obtained in this work also shed light on a previously unexplained aspect of volatility behaviour observed in a number of large static chambers whereby an active enhancement of OH chemistry (Salo et al., 2011) induced a rapid increase in the aerosol's volatility, O:C ratio, and mass (Donahue et al., 2012). It was initially proposed that these outcomes were due to the condensation of highly oxygenated volatile material as a result of reactions with OH. Our PAM chamber results support this hypothesis because the subsequent OH-induced oxidation of the continuously replenished mixture of ozonolysis products yielded a near-constant volatility regardless of the level of OH exposure in the PAM chamber, and the addition of more volatile products formed by OH oxidation compensates for the oxidative ageing effect of the ozonolysis products.

As can be seen in Figure 5b, the $S_{VFR}$ of the SOA generally increased with increasing OH exposure. The change in $S_{VFR}$ for the LIM-O3 case is small compared to that for the other two terpenes. Figure 6 summarises the key results of this work and shows that the change in $S_{VFR}$ is related to a higher evaporation temperature for the least volatile aerosol material ($T_{VFR0.1}$) in the case of $\alpha$- and $\beta$-pinene, but that the relationship between $T_{VFR0.5}$ and $S_{VFR}$ is very weak or non-existent. This suggests that the formation of ELVOCs will not significantly influence the general volatility, but does clearly broaden the volatility distribution. To compare the ageing-induced changes in $S_{VFR}$ to the derived ageing parametrisation of Emanuelsson et al. (2014), a linear fit of the $S_{VFR}$ to the OH concentration was constructed. The derived linear relationship predicted $S_{VFR}$ increases of $3.8 \pm 0.1\%$ ($R^2$=0.99) and $3.2 \pm 0.2\%$ ($R^2$=0.98) for $\alpha$- and $\beta$-pinene, respectively, upon exposure to an OH radical concentration of $1 \times 10^6$ $cm^{-3}$ for one hour. This observed increase in $S_{VFR}$ stands in contrast to the observations from the SAPHIR experiments (Emanuelsson et al., 2014), in which ageing reduced aerosol complexity (-0.9±0.3%), and highlights the differences between the processes that are important in larger static chambers like SAPHIR and those that are most relevant in the PAM chamber. Particularly, the SAPHIR and PAM chambers differ in terms of reaction times and dilution; the longer reaction times and successive dilutions in SAPHIR experiments allow the products that are formed to be redistributed between the gas and the particle phases, with the gas phase compounds subsequently being oxidised and removed over time. Conversely, in the PAM chamber there will always be an inflow of more volatile fresh material such that the overall volatility remains constant but the proportion of low volatility compounds gradually increases, broadening the volatility distribution (and thus increasing $S_{VFR}$) as the OH exposure increases. In the atmosphere, both conditions are likely to obtain to at least some extent; successive dilutions will occur above forest canopies, but there will also be a constant inflow of fresh material depending on the extension of the forest region. Therefore, data from both PAM chambers and larger static chambers such as SAPHIR are needed to properly describe the evolution of the thermal properties of atmospheric SOA.

## 5. Conclusions

We have investigated the thermal properties of SOA produced from three terpenes by ozonolysis and oxidation initiated by OH radicals, and studied the effects of ageing due to OH exposure on these SOA. The SOA formed by ozonolysis were less volatile than those formed by OH radical oxidation. Additionally, the thermal properties of SOA formed from limonene clearly differed from those of SOA derived from α- and β-pinene, especially when oxidation was achieved by ozonolysis. In general, the volatility distribution of the terpene-derived SOA was broadened when they were exposed to OH radicals, and the degree of broadening increased with the radical concentration. The evaporation temperature of the aerosols' least volatile fractions also increased with the degree of OH exposure, demonstrating that OH chemistry has important effects on the formation of low volatility organic compounds. However, in contrast to the results obtained in static chamber experiments, the overall volatility did not increase with increasing OH exposure. These findings reveal that there are two important opposing effects that influence the evolution of volatility upon ageing in the atmosphere, and shows that results from both flow reactors and static chambers are needed to understand the development of the thermal properties of ambient SOA.

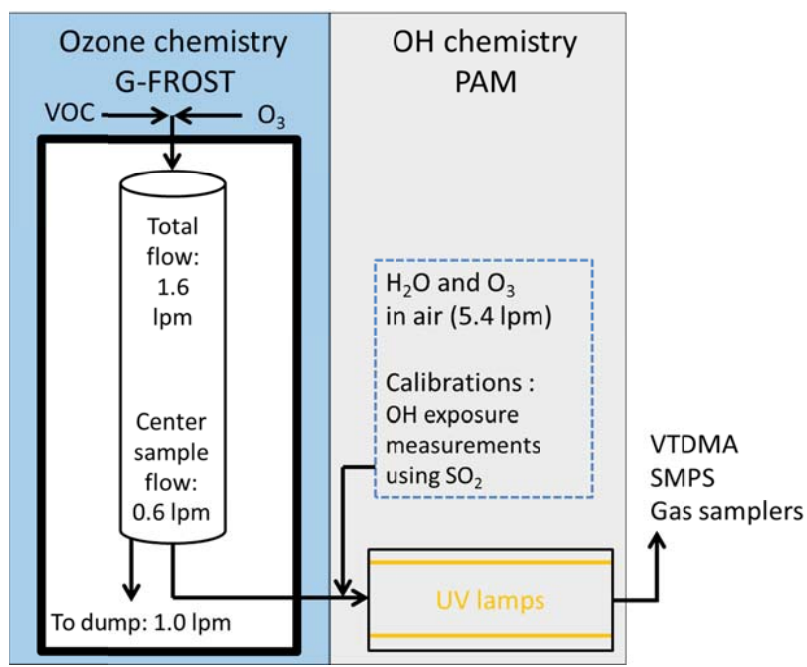

Figure 1. Schematic of the experimental set-up. By selective production of oxidants in G-FROST and the PAM chamber four different oxidation conditions were evaluated: OH-dominated oxidation (only PAM); ozone-dominated oxidation with some influence from OH radicals (G-FROST); ozone-dominated oxidation (G-FROST with added 2-butanol); and sequential ozone and OH oxidation (G-FROST+PAM).

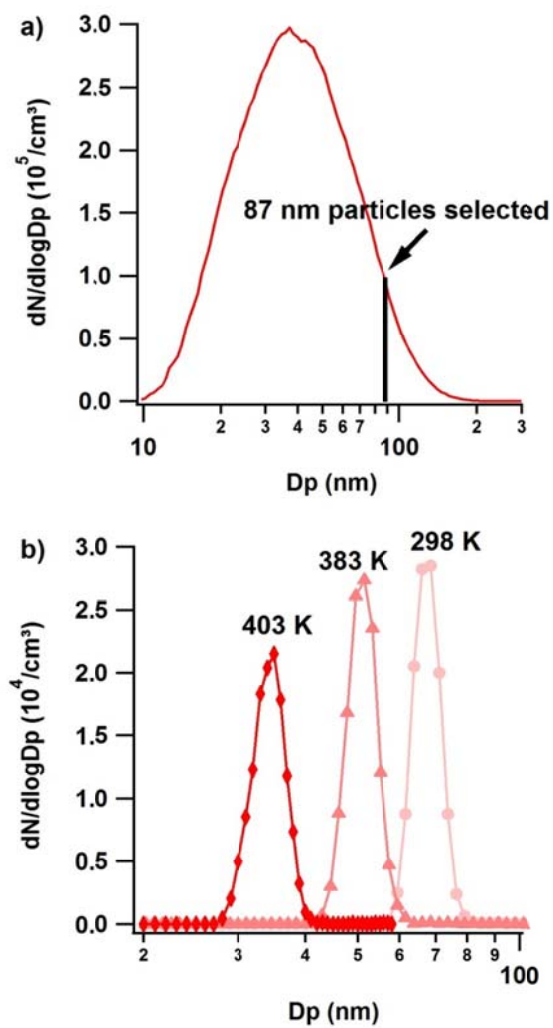

Figure 2: Example of the VTDMA method. (a) Initial number size distribution. (b) residual number size distributions after selection with the first DMA and evaporation in the oven unit at three temperatures (298, 383, 403 K).

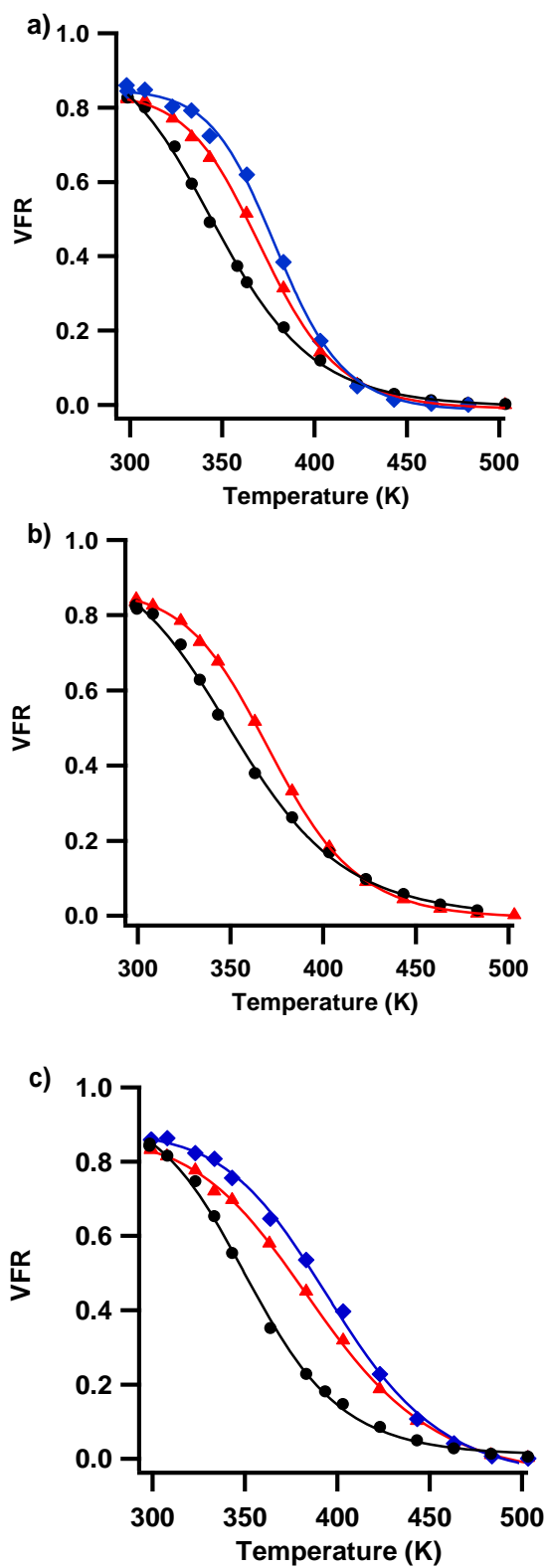

Figure 3 a-c: The VFR as a function of the evaporation temperature for α-pinene (a), β-pinene (b) and limonene (c) in the O3- (blue), MIX- (red) and OH-experiments (black). The solid lines represent the fitted function (equation) used to derive T$_{position}$ and S$_{VFR}$ for each experiment.

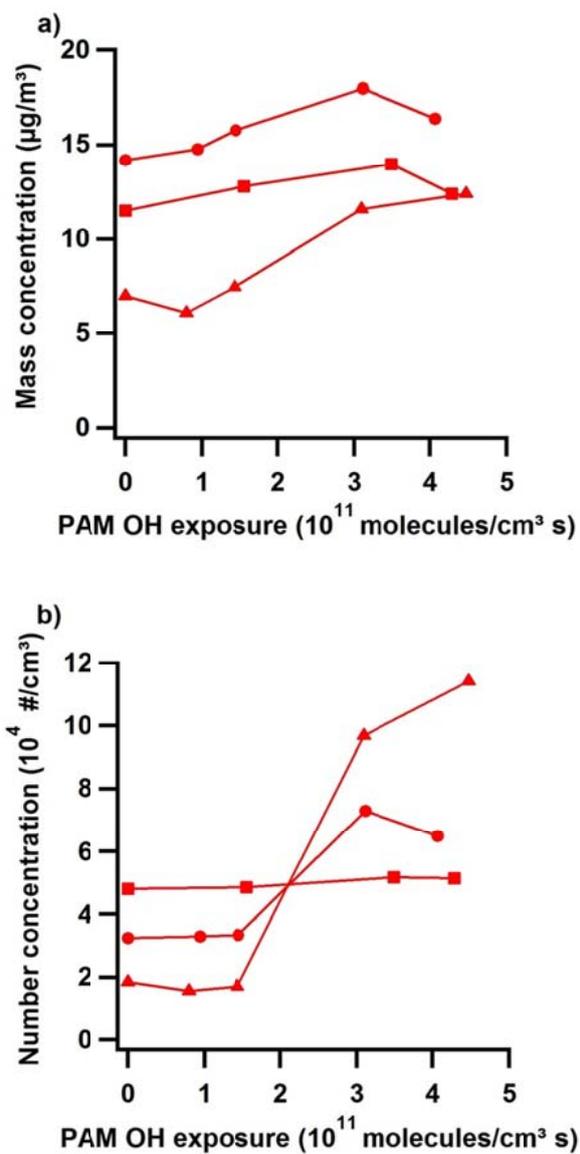

Figure 4: Particle mass (a) and number (b) concentrations as a function of PAM OH-exposure for α-pinene (circles), β-pinene (triangles) and limonene (squares) during sequential oxidation (SEQ).

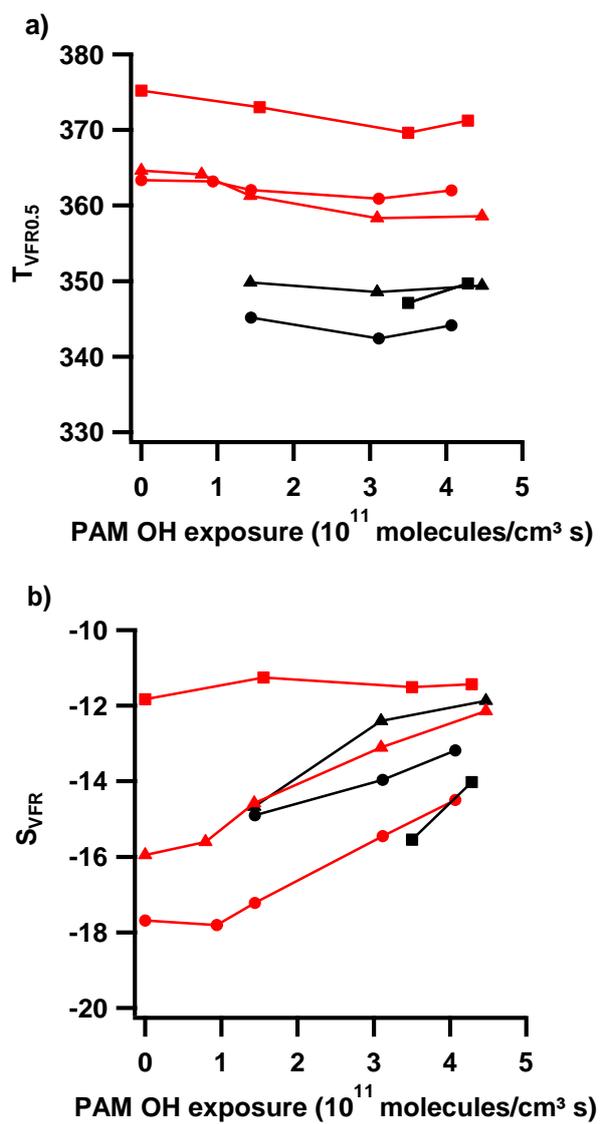

Figure 5: $T_{VFR0.5}$ (a) and $S_{VFR}$ (b) as a function of OH-exposure for α-pinene (circles), β-pinene (triangles) and limonene (squares) for the OH- (black) and SEQ- experiments (red).

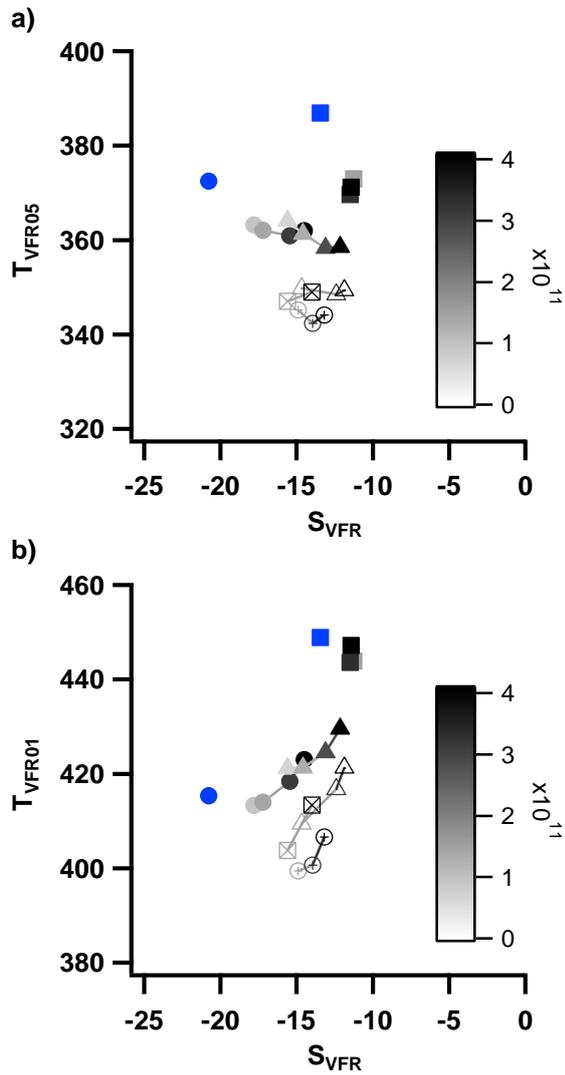

Figure 6: $T_{VFR0.5}$ (a) and $T_{VFR0.1}$ (b) plotted as functions of $S_{VFR}$ for α-pinene (circles), β-pinene (triangles) and limonene (squares) in the O3 (blue), SEQ (closed) and OH (open) experiments. Darker markers correspond to higher levels of OH exposure.

Table 1: Particle mass, number and volatility parameters for the experiments. The statistical errors on the averages are given as one standard deviation for consecutive measurements. b.d. : below limit of detection

| Exp. | Terpene | Mass ($\mu$g m$^{-3}$) | Number ($10^4$ cm$^{-3}$) | $T_{VFR0.5}$ (K) | $T_{VFR0.1}$ (K) | $S_{VFR}$ |
|------|---------|------|--------|---------|---------|------|
| **α-O3** | α-pinene | 9.1±0.9 | 0.4±0.05 | 372.5±0.1 | 415.4±0.1 | -20.8±1.3 |
| **α-MIX** | α-pinene | 19.9±0.6 | 3.2±0.04 | 363.2±0.1 | 413.5±0.1 | -17.7±0.8 |
| **α-SEQ** | α-pinene | 20.7±0.4-25.2±0.8 | 3.3±0.04-7.3±0.8 | 360.9±0.1-363.2±0.1 | 414.0±0.1-423.0±0.1 | (-17.8±0.6)-(-14.5±0.4) |
| **α-OH** | α-pinene | 2.8±0.1-21.8±0.7 | 3.0±0.1-17.4±0.8 | 342.4±0.2-345.2±0.2 | 399.5±0.2-406.6±0.2 | (-14.9±1.0)-(-13.2±0.6) |
| **β-O3** | β-pinene | b.d. | b.d. | b.d. | b.d. | b.d. |
| **β-MIX** | β-pinene | 9.8±0.1 | 1.9±0.02 | 364.6±0.1 | 421.3±0.1 | -16.0±0.2 |
| **β-SEQ** | β-pinene | 8.5±0.1-17.4±0.2 | 1.6±0.02-11.4±0.4 | 358.3±0.1-364.1±0.1 | 421.3±0.1-424.6±0.1 | (-15.6±0.3)-(-12.4±0.4) |
| **β-OH** | β-pinene | b.d.-13.7±0.2 | 2.9±0.1-11.2±0.3 | 348.5±0.1-349.8±0.1 | 409.5±0.1-421.3±0.2 | (-14.7±0.3)-(-11.9±0.7) |
| **L-O3** | Limonene | 9.5±0.1 | 3.3±0.02 | 386.9±0.1 | 448.9±0.1 | -13.5±0.8 |
| **L-MIX** | Limonene | 10.9±0.1 | 5.8±0.1 | 375.2±0.1 | 445±0.1 | -11.8±0.5 |
| **L-SEQ** | Limonene | 16.1±0.2-19.2±0.4 | 4.9±0.02-5.2±0.1 | 369.6±0.1-373.0±0.1 | 443.9±0.2-447.2±0.1 | (-11.5±0.6) -(-11.3±0.4) |
| **L-OH** | Limonene | 22.1±0.3-26.7±0.6 | 22.8±1.1-31.4±1.4 | 347.1±0.1-349.7±0.1 | 403.8±0.1-413.4±0.1 | (-15.6±0.7)-(-14.0±0.7) |

**Acknowledgments**

The research presented is a contribution to the Swedish strategic research area ModElling the Regional and Global Earth system, MERGE. This work was supported by Formas (grant numbers 214-2010-1756, 214-2013-1430); the Swedish Research Council (grant number 2014-05332)